\documentclass[11pt]{lmcs}
\usepackage{latexsym,amsmath,amssymb,amsthm,stmaryrd,url}
\usepackage{mathrsfs}
\usepackage{graphicx}
\usepackage{caption}
\usepackage{float}
\usepackage{subcaption}
\usepackage{diagbox}
\usetikzlibrary{arrows,shapes,chains}

\newcommand{\dda}{\mathord{\mbox{\makebox[0pt][l]{\raisebox{-.4ex}{$\downarrow$}}$\downarrow$}}}
\newcommand{\dua}{\mathord{\mbox{\makebox[0pt][l]{\raisebox{.4ex}{$\uparrow$}}$\uparrow$}}}

\newcommand{\da}{\downarrow\!\!}
\newcommand{\Da}{\Downarrow\!}
\newcommand{\ra}{\rightarrow\!}
\newcommand{\ua}{\uparrow\!\!}%


\begin{document}

\title[Some Consistent Power Constructions]{Some Consistent Power Constructions}

\author[C.~Zhou]{Chengyu Zhou \lmcsorcid{0009-0005-8021-8863}}[a]
\author[Q.~Li]{Qingguo Li \lmcsorcid{0000-0002-4314-7489}}[b]

\address{School of Mathematics, Hunan University, Changsha, Hunan, 410082, China}
\email{zhouchy126@126.com}

\address{School of Mathematics, Hunan University, Changsha, Hunan, 410082, China}
\email{liqingguoli@aliyun.com (Corresponding author)}

\begin{abstract}
Consistent Hoare, Smyth and Plotkin power domains are introduced and discussed by Yuan and Kou.
 The consistent algebraic operation $+$ defined by them is a binary partial Scott continuous operation satisfying the requirement: $a+b$ exists whenever there exists a $c$  which is greater than $a$ and $b$.
 We extend the consistency to be a categorical concept and obtain an approach to generating consistent monads from monads on dcpos whose images equipped with some algebraic operations.
 Then we provide two new power constructions over domains: the consistent Plotkin index power domain and the consistent probabilistic power domain.
 Moreover, we verify these power constructions are free.
\end{abstract}

\maketitle

\section{Introduction}
Domain, presented for modeling functional programming languages by Dana Scott, is a kind of abstract mathematical structure of data types.
 In the 1970s, Plotikin, Smyth, and etc. gave three power constructions over domains: Smyth power domain, Plotkin power domain and Hoare power domain, for denoting non-determinism computation. The classical non-determinism generator $``or"$ was described as an algebraic operation $+$ which is idempotent, associative and commutative (\cite{Hennessy1979}).
 In \cite{Moggi1991}, Moggi proved that the power constructions are some special monads on domains, and showed these monads are computable in the $\lambda_c$-calculus, a programming language presented by himself.
 In 1990, Jones developed Lawson's valuation (\cite{Lawson1982}) to be a monad on dcpos for depicting probabilistic choices.
 She called this monad a probabilistic power and examined that it also satisfies the requirements of Moggi's programming language (\cite{Jones1990}).
 A probabilistic choice states that a program could go to a state with probability $r$ and go to another with probability $(1-r)$ for some real number $r\in[0,1]$.
 Moreover, a probabilistic choice could be described as an algebraic operation with some properties.
 Based on the research of Jones, many power constructions have been found such as index power, mixed power and so on (\cite{Varacca2003,Klaus2017,Jia2021}).

In  \cite{Yuan2014,Yuan2014a,Yuan2014b}, Yuan and Kou discussed consistent Smyth, Hoare, Plotkin power construcions over domains.
 They set consistent semilattices, consistent inflationary semilattices and consistent deflationary semilattices as dcpos which carry a partial Scott continuous semilattice operation $+$ obeying different laws and satisfying the consistent condition: $a+b$ exists if there is a $c$ such that $a,b\leq c$.
 Then they defined three consistent powers as free constructions from the category of domains to those of consistent semilattices, consistent inflationary semilattices and consistent deflationary semilattices.

In this paper, we are interested in finding out if there are more consistent power constructions.
 It is hard to certify whether every forgetful functor from the category of dcpos with consistent algebraic operations to that of dcpos has a left adjoint.
 However, each free dcpo-algebras gives us a monad having some algebraic features on dcpos.
 Thus, we have some results in Section \ref{SecMonad}.
 In Subsection \ref{D-AlgebraicTheory}, we extract the consistent condition to be a functor from $\mathbf{POS}$ to $\mathbf{SET}$.
 Then, in Section \ref{SecMonad}, we present an approach to revising a monad on dcpos which has some algebraic features to a monad having some consistent algebraic features by transfinite inductions and the pursuit of a sub construction named sub o.b.d.algebra.

 In Section \ref{SecIndex}, we provide a subset of a Plotkin index power domain, and we prove that the Scott closure of the subset with the induced order gives the free construction from the category of domains to that of consistent dcpo quasi-cones.

 In Section \ref{SecKeg}, we discuss the consistent probabilistic power domain. The consistent probabilistic power domain, which gives a free construction from the category of domains to that of consistent kegespitzen, is characterized as the minimum subdcpo of the probabilistic power domain that contains these simple valuations lower than some Dirac valuations.

\section{Preliminary}

\subsection{Domain theory}
Let $P$ be a poset.
 An element $x\in P$ is an upper bound (resp., a lower bound) of a subset $S$ iff $x\geq y$ (resp., $x\leq y$) for all $y\in S$.
 The set of upper bounds (resp., lower bounds) of $S$ is denoted as $S^u$ (resp., $S^l$).
 For a subset $A$ of $P$, let $\da A=\{y:\exists x\in A \text{ s.t. } y\leq x\}$ and $\ua A=\{y:\exists x\in A \text{ s.t. } y\geq x\}$.
 The subset $A$ is a lower subset (resp., an upper subset) iff $A=\da A$ (resp., $A=\ua A$).
 For every element $x$ of $P$, we write $\da x$ (resp., $\ua x$) for $\da\{x\}$ (resp., $\ua x$).
 A subset $D$ is directed iff it is non-empty and every $x$ and $y$ of $D$ has an upper bound in $D$.
 A subset $D'$ is co-final in the $D$ iff $D'\subseteq D$ and each $x\in D$ has an upper bound in $D'$, and we say that $D'$ is a co-final subset of $D$.
 If every directed subset of $P$ has a sup, then $P$ is a directed complete poset (a dcpo for short).
 In the poset $P$, we say that $y$ is way-below $x$ in symbols $y\ll x$ iff $x\leq\bigsqcup D$ could imply there is a $d\in D$ with $y\leq d$ for every directed subset $D$, where $\bigsqcup D$ is a sup of $D$.
 we write $\dda x$ for the set of these elements are way-below $x$.
 If $\dda x$ forms a directed subset and the sup of $\dda x$ is $x$, then we say that the poset $P$ is continuous.
 Each continuous dcpo is called a domain.
 A subset $B$ of $L$ is a basis of $L$ iff for every $x\in L$, $\dda x\cap B$ is directed and has $x$ as a sup.
 \begin{rems}
   Let $B$ be a subset of poset $P$.
    The following statements hold.
   \begin{enumerate}[(i)]
     \item If $S_x$ is a directed subset of $\dda x\cap B$ and it has $x$ as a sup for each $x$, then $B$ is a basis.
     \item If $B$ is a basis, then each co-final subset of $\dda x\cap B$ is a directed subset which has a sup $x$.
     \item If $B$ is a basis, each directed subset of $\dda x\cap B$ with a sup $x$ is co-final in $\dda x\cap B$.
     \item If $B$ is a basis, $D$ is a directed subset of $P$ and $S_d$ is co-final subset of $\dda d\cap B$ for all $d\in D$, then $\bigcup_{d\in D} S_d$ is co-final in $\dda\bigsqcup D\cap B$.
     \item The poset $P$ is continuous iff it has a basis.
   \end{enumerate}
 \end{rems}

A map $f:P\ra Q$ between posets is order-preserving iff $a\leq b$ implies $f(a)\leq f(b)$.
 It is order-embedding iff it preserves the order and $f(a)\leq f(b)$ implies $a\leq b$.
 It is Scott continuous iff $f(\bigsqcup D)=\bigsqcup f(D)$ holds for every directed subset $D$ which has a sup.
 The category $\mathbf{POS}$ is the category of posets and order-preserving maps, the category $\mathbf{DCPO}$ (resp., $\mathbf{DOM}$) is the category of dcpos (resp., domains) and Scott continuous maps.

We use the notation $\prod_{i=1}^{n}$ for $n$-array products in some categories.
 If every object (resp., morphism) $A_i$ coincides with an object (resp., a morphism) $A$, then we write $\prod^{n}A$ instead of $\prod_{i=1}^{n}A_i$.
 The product $\prod_{i=1}^{n}P_i$ of objects in $\mathbf{POS}$ is provided by $\{(p_1,p_2,\ldots,p_n):p_i\in P_i\}$ with the pointwise order, the product $\prod_{i=1}^nf_i:\prod_{i=1}^{n}P_i\ra\prod_{i=1}^{n}Q_i$ of morphisms $f_i:P_i\ra Q_i$ in $\mathbf{POS}$ is given by $\prod_{i=1}^nf_i((p_1,p_2,\ldots,p_n))=(f_1(p_1),f_2(p_2),\ldots,f_n(p_n))$; so are $\mathbf{DCPO},\mathbf{DOM}$.

If a lower subset of a poset is closed for directed sups, then it is a Scott closed subset.
 The complements of Scott closed subsets are Scott opens.
 All Scott opens form a topology, called Scott topology.

 \begin{defi}
    An abstract basis is a set $B$ endowed with a binary relation $\prec$ which is transitive and has the following finite interpolation property:
    \begin{center}
      for every finite subset $F$ and every element $z\in B$, one has\par
      $F\prec z\Rightarrow(\exists y\in B)F\prec y\prec z$.
    \end{center}
     Round ideals are lower sets which are directed w.r.t. relation $\prec$.
     The round ideal completion $RId(B,\prec)$ of $B$ is the set of round ideals ordered by inclusion.
 \end{defi}
  \begin{prop}
  \label{AbsBasisPPT}
    Let $(B,\prec)$ be an abstract basis.
     Then the following statements hold:
    \begin{enumerate}[(i)]
      \item $RId(B)$ is a domain,
      \item for each $x\in B$, $\da_\prec x:=\{y:y\prec x\}$ is a round ideal,
      \item each round ideal $R$ is a sup of $\{\da_\prec x:x\in R\}$,
      \item $R_1\ll R_2$ in $RId(B,\prec)$ iff there is an $x\in R_2$ such that $R_1\subseteq\da_\prec x$,
      \item $\da_\prec x\ll R$ for each $x\in R$.
    \end{enumerate}
  \end{prop}

\subsection{\texorpdfstring{$D$-algebraic theory and order bound system}{D-algebraic theory and order bound system}}
\label{D-AlgebraicTheory}
In this subsection, we give some new notions to be used in this paper.
 Most of them are the variants of some familiar notions.

\begin{defi}
  A $D$-algebraic theory $\mathbb{T}$ is a tuple $(\Sigma_0,\Sigma_1,\lvert\cdot\rvert,M,\theta,\Lambda)$, where:
  \begin{enumerate}[(i)]
    \item $\Sigma_0$ and $\Sigma_1$ are the sets of given operators, and $\Sigma_0\cap\Sigma_1=\emptyset$,
    \item $\lvert\cdot\rvert:\Sigma_0\cup\Sigma_1\ra\mathbb{N}$ is a map,
    \item $M$ is a set of given dcpos,
    \item $\theta:A\ra M\cup\{*\}$ is a map, for which $A=\{(g,n)\in\Sigma_1\times\mathbb{N}:n\leq\lvert g\rvert\}$,
    \item $\Lambda$ is a set of laws in form of $e_1\sqsubseteq e_2$ or $e_1=e_2$, for which $e_1,e_2$ are the expressions formed from a convenient stock of variables by applying the given operators of $\Sigma_0\cup\Sigma_1$.
  \end{enumerate}
\end{defi}

\begin{exa}
  Let
  \begin{enumerate}[(i)]
    \item $\Sigma_0=\{+,\underline{0}\}$,
    \item $\Sigma_1=\{\star\}$,
    \item $\lvert +\rvert=2$, $\lvert \underline{0}\rvert=0$, and $\lvert\star\rvert=2$,
    \item $M=\{\overline{\mathbb{R}}_{\geq 0}\}$, where $\overline{\mathbb{R}}_{\geq 0}$ is the set of the extended non-negative real numbers endowed with the usual order,
    \item $\theta(\star,1)=\overline{\mathbb{R}}_{\geq 0}$ and $\theta(\star,2)=*$,
    \item $\Lambda$ be consisted of the following laws:
    \begin{align}
    &(1)\quad x+ y=y+ x                    & \hspace{1em} &(5)\quad 0\star x=\underline{0} \nonumber \\
    &(2)\quad (x+ y)+ z=x+(y+ z) &              &(6)\quad (rs)\star x=r\star(s\star x) \nonumber\\
    &(3)\quad \underline{0}+ x=x                &              &(7)\quad r\star(x+y)=r\star x+ r\star y \nonumber\\
    &(4)\quad 1\star x=x                             &              &(8)\quad (r+s)\star x=r\star x+ s\star x. \nonumber
   \end{align}
  \end{enumerate}
   Then $(\Sigma_0,\Sigma_1,\lvert\cdot\rvert,M,\theta,\Lambda)$ is a $D$-algebraic theory.
    We denote it as $\mathbb{C}$ and we call it the dcpo cone theory.
\end{exa}

In what follows, we call the operation $\star$ scalar multiplication and simplify $r\star x$ as $rx$, the operation $+$ is called addition.

\begin{defi}
  Let $\mathbb{T}=(\Sigma_0,\Sigma_1,\lvert\cdot\rvert,M,\theta,\Lambda)$ be a $D$-algebraic theory. A dcpo $\mathbb{T}$-algebra is a pair $(L,i)$, where $L$ is a carrier dcpo and $i$ is an interpretation satisfying:
  \begin{enumerate}[(i)]
    \item for each $f\in\Sigma_0$, $i(f)$ is a Scott continuous map from the product $\prod^{\lvert f\rvert}L$ to $L$,
    \item for each $m\in M$, $i(m)=m$,
    \item $i(*)=L$,
    \item for each $g\in\Sigma_1$, $i(g):i(\theta((g,1)))\times i(\theta((g,2)))\times\cdots\times i(\theta((g,\lvert g\rvert)))\ra L$ is a Scott continuous map,
    \item \textbf{obeying the laws}: the equation or inequation of interpreted operators corresponding to each law is true.
  \end{enumerate}
\end{defi}

We speak of $i(f)$ as the operation of $f$ on $L$.

\begin{defi}
  A homomorphism $h$ between dcpo $\mathbb{T}$-algebras $(X,i)$ and $(Y,j)$ is a Scott continuous map from $X$ to $Y$ and preserves each operation:
  \begin{enumerate}[(i)]
    \item for each $f\in\Sigma_0$, $h(i(f)(x_1,x_2,\ldots,x_{\lvert f\rvert}))=j(f)(h(x_1),h(x_2),\ldots,h(x_{\lvert f\rvert}))$,
    \item for each $g\in\Sigma_1$, $h(i(g)(x_1,x_2,\ldots,x_{\lvert f\rvert}))=j(g)(h_1(x_1),h_2(x_2),\ldots,h_{\lvert f\rvert}(x_{\lvert f\rvert}))$, where $h_i=h$ if $\theta(g,i)=*$ and $h_i$ is the identity of $i(\theta((g,i)))$ if $\theta((g,i))\neq*$.
  \end{enumerate}
\end{defi}

We denote by $\mathbf{DCPO}\mathbb{T}$ the category whose objects are dcpo $\mathbb{T}$-algebras and morphisms are homomorphisms.
Moreover, we denote the full subcategory whose objects are domain $\mathbb{T}$-algebras (that is, carriers are domains) by $\mathbf{DOM}\mathbb{T}$.

\begin{defi}
  Let $\mathbb{T}=(\Sigma_0,\Sigma_1,\lvert\cdot\rvert,M,\theta,\Lambda)$ be a $D$-algebraic theory.
  An order bound system $\Omega$ of $\mathbb{T}$ is a functor from $\Sigma_0$ to the co-product category $\coprod_{f\in\Sigma_0}\mathbf{A}_f$ such that $\Omega f$ is in $\mathbf{A}_f$ for each $f$, where each $\mathbf{A}_f$ is a category whose objects are the functors $F:\mathbf{POS}\ra\mathbf{SET}$ satisfying:
   \begin{enumerate}[(i)]
     \item if $P$ is a poset, then $FP$ is a subset of $\prod^{\lvert f\rvert}\!P$,
     \item if $h:P\ra Q$ is an order-preserving map and $(x_1,x_2,\ldots,x_{\lvert f\rvert})\in FP$, then
     \begin{center}
       $(h(x_1),h(x_2),\ldots,h(x_{\lvert f\rvert}))\in FQ$,
     \end{center}
     that is $\prod^{\lvert f\rvert}\!h(FP)\subseteq FQ$,
     \item if $h:P\ra Q$ is an order-preserving map, then $Fh$ is equal to $(\prod^{\lvert f\rvert}\!h)|^{FQ}_{FP}$, the restriction and co-restriction of $\prod^{\lvert f\rvert}\!h$;
   \end{enumerate}
   and whose morphisms are natural isomorphisms.
   In general, we call $\Omega f$ an order bound of $f$ and write it as $\Omega^f$.
\end{defi}

\begin{rem}
\label{UoDFuctors}
  Let $f$ be an operator of $\Sigma_0$ of a $D$-algebraic theory $\mathbb{T}$, $\prod^{\lvert f\rvert}:\mathbf{POS}\ra\mathbf{POS}$ the $\lvert f\rvert$-array product functor that maps a poset $P$ to $\prod^{\lvert f\rvert} P$ and a order-preserving map $g$ to $\prod^{\lvert f\rvert}g$, $U:\mathbf{POS}\ra \mathbf{SET}$ the forgetful functor.
   Then it is easy to verify that the composition $U\circ\prod^{\lvert f\rvert}$ is in $\mathbf{A}_f$.
\end{rem}

\begin{defi}
\label{Obda}
  Let $\mathbb{T}=(\Sigma_0,\Sigma_1,\lvert\cdot\rvert,M,\theta,\Lambda)$ be a $D$-algebraic theory and $\Omega$ an order bound system of $\mathbb{T}$.
   We denote by $\mathbf{DCPO}(\mathbb{T},\Omega)$ (resp., $\mathbf{DOM}(\mathbb{T},\Omega)$) the category whose objects are pairs $(L,i)$, called an order bound dcpo $\mathbb{T}$-algebra (resp., an order bound domain $\mathbb{T}$-algebra), consisting of  a carrier dcpo (resp., a carrier domain) $L$ and an interpretation $i$ such that:
   \begin{enumerate}[(i)]
     \item for each $m\in M$, $i(m)=m$,
     \item $i(*)=L$,
     \item for each $f\in\Sigma_0$ and $(x_1,x_2,\ldots,x_n)\in\Omega^fL$, $i(f)(x_1,x_2,\ldots,x_n)$ exists in $L$. Moreover, if $D$ is a directed subset of $\Omega^fL$ and $\bigsqcup D\in\Omega^fL$, then $\bigsqcup i(f)(D)=i(f)(\bigsqcup D)$, (whenever an operation satisfies this condition, we say that it is partial Scott continuous.)
     \item for each $g\in\Sigma_1$, $i(g):i(\theta((g,1)))\times i(\theta((g,2)))\times\cdots\times i(\theta((g,\lvert g\rvert)))\ra L$ is a Scott continuous map,
     \item \textbf{skipped obeying the laws}: each equation or inequation, deduced by laws and the property that operations preserve the order, is true, whenever all occurrences of $i(f)$ in the equation or inequation are acting on the elements of $\Omega^fL$ for all $f\in\Sigma_0$, 
   \end{enumerate}
   and whose morphisms are those Scott continuous maps $h:(X,i)\ra (Y,j)$ such that
   \begin{enumerate}[(i)]
     \item[(vi)] if $f\in\Sigma_0$ and $(x_1,x_2,\ldots,x_n)\in\Omega^fX$, then
     \begin{center}
     $h(i(f)(x_1,x_2,\ldots,x_n))=j(f)(h(x_1),h(x_2),\ldots,h(x_n))$,
   \end{center}
     \item[(vii)] $h$ preserves $i(g)$ for each $g\in\Sigma_1$.
   \end{enumerate}
   We also call a morphism of $\mathbf{DCPO}(\mathbb{T},\Omega)$ a homomorphism.
\end{defi}
\begin{rem}
     For an operator $f$ of $\Sigma_0$ of a $D$-algebraic theory $\mathbb{T}$ and an order bound system $\Omega$ of $\mathbb{T}$, if $\Omega^f=U\circ\prod^{\lvert f\rvert}$ which is the composition of functors given in Remark \ref{UoDFuctors}, then the operation $i(f)$ of each object of $\mathbf{DCPO}(\mathbb{T},\Omega)$ (resp., $\mathbf{DOM}(\mathbb{T},\Omega)$) is not a partial map, and then the category $\mathbf{DCPO}(\mathbb{T},\Omega)$ (resp., $\mathbf{DOM}(\mathbb{T},\Omega)$) is equivalent to  $\mathbf{DCPO}\mathbb{T}$ (resp., $\mathbf{DOM}\mathbb{T}$).
\end{rem}
\begin{exa}
   Let
   \begin{enumerate}[(i)]
     \item $\Omega^+_1$ be a functor for which $\Omega^+_1P=\{(x,y):\{x,y\}^u\neq\emptyset\}$,
     \item $\Omega^+_2$ be a functor for which $\Omega^+_2P=\{(x,y):\{x,y\}^l\neq\emptyset\}$,
     \item $\Omega^+_3$ be equal to $U\circ\prod^2$.
   \end{enumerate}
    Then each $\mathbf{DCPO}(\mathbb{C},\{\Omega^+_i\})$ is different from others, and $\mathbf{DCPO}(\mathbb{C},\{\Omega^+_3\})$ is equivalent to $\mathbf{DCPO}\mathbb{C}$.
\end{exa}

\section{Modify monads on dcpos}
\label{SecMonad}
Let $\mathbb{T}=(\Sigma_0,\Sigma_1,\lvert\cdot\rvert,M,\theta,\Lambda)$ be an $D$-algebraic theory, $\Omega$ an order bound system of $\mathbb{T}$, $(L,i)$ be in $\mathbf{DCPO}(\mathbb{T},\Omega)$ and $S$ a subdcpo of $L$.
  For each $g\in\Sigma_1$, we obtain a subdcpo of $i(\theta((g,1)))\times i(\theta((g,2)))\times\cdots\times i(\theta((g,\lvert g\rvert)))$ by restricting variables ranged in $S$, and denote it as $E_{g}(S)$.
  For example, if $\mathbb{T}$ is the dcpo cone theory $\mathbb{C}$, then $E_\star(S)=\overline{\mathbb{R}}_{\geq 0}\times S$.

\begin{defi}
  We say that $S$ is a sub order bound dcpo algebra of $(L,i)$ (a sub o.b.d.algebra for short) iff $i(f)(\Omega^fS)\subseteq S$ for all $f\in\Sigma_0$ and $i(g)(E_g(S))\subseteq S$ for all $g\in\Sigma_1$.
\end{defi}

\begin{rem}
\label{SubObdaIsObda}
  If $S$ is a sub o.b.d.algebra of $(L,i)$, then $(S,i')$ is in $\mathbf{DCPO}(\mathbb{T},\Omega)$, where $i'$ is the interpretation such that:
  \begin{enumerate}[(i)]
    \item $i'(m)=m$ for all $m\in M$,
    \item $i'(*)=S$,
    \item $i'(f)$ and $i'(g)$ are operations induced by $i(f)$ and $i(g)$, respectively.
  \end{enumerate}
\end{rem}

\begin{prop}
  Each non-empty intersection of sub o.b.d.algebras is a sub o.b.d.algebra.
\end{prop}
\proof
  Assume $\{S_j\}_{i\in J}$ is a non-empty family of sub o.b.d.algebras of $(L,i)$, and $f\in\Sigma_0$.
   Obviously, $\bigcap_{j\in J}S_j$ is a subdcpo. For each $j\in J$, the inclusion $inc_j: \bigcap_{k\in J}S_k\ra S_j$ preserves the order.
   Then $\Omega^f(\bigcap_{j\in J}S_j)\subseteq \Omega^fS_j$.
   Hence, $i(f)(\Omega^f(\bigcap_{k\in J}S_k))$ is contained in $S_j$ for every $j\in J$.
   It implies that $i(f)(\Omega^f(\bigcap_{j\in J}S_j))\subseteq\bigcap_{j\in J}S_j$.
   It is easy to know that $\bigcap_{j\in J}S_j$ is closed for the operation $i(g)$ for each $g\in\Sigma_1$.
   Thus $\bigcap_{j\in J}S_j$ is a sub order bounded dcpo algebra.
\qed

We write the minimum sub o.b.d.algebra containing $S$ as $cl^{(\mathbb{T},\Omega)}_{(L,i)}(S)$ and call it the sub o.b.d.algebra closure of $S$.
 Let $\mathcal{D}_s(B)$ be the set of the directed sups of all directed subset of $B$, where $B$ is a subset of $L$.
 Then we set:
 \begin{align}
     &S^0=S,\nonumber\\
     &S^{\alpha}_0=\bigcup_{\beta<\alpha}S^{\beta},\nonumber\\
     &S^{\alpha}_1=\bigcup_{g\in\Sigma_1}i(g)(E_g(S^{\alpha}_0)),\nonumber\\
     &S^{\alpha}_2=\bigcup_{f\in\Sigma_0}i(f)(\Omega^fS^{\alpha}_0),\nonumber\\
     &S^{\alpha}=\mathcal{D}_s(S^{\alpha}_0\cup S^{\alpha}_1\cup S^{\alpha}_2),\nonumber\\
     &S^*=\bigcup_{\alpha\in ORD}S^{\alpha}.\nonumber
 \end{align}

\begin{prop}
\label{GenSC}
  $S^*=cl^{(\mathbb{T} ,\Omega)}_{(L,i)}(S)$.
\end{prop}
\proof
  Obviously, $cl^{(\mathbb{T} ,\Omega)}_{(L,i)}(S)\supseteq S^*$ since $(cl^{(\mathbb{T} ,\Omega)}_{(L,i)}(S))^\alpha=cl^{(\mathbb{T} ,\Omega)}_{(L,i)}(S)$ for all ordinal number $\alpha$.

  For the converse, we show that $cl^{(\mathbb{T} ,\Omega)}_{(L,i)}(S)\subseteq S^*$ by verifying that $S^*$ is a sub o.b.d.algebra.
   Because the induction is limited by $cl^{(\mathbb{T} ,\Omega)}_{(L,i)}(S)$, there is an ordinal number $\alpha$ such that $S^\alpha=S^{\alpha+1}=S^*$.
   For each $f\in\Sigma_0$ and $(x_1,\ldots,x_{\lvert f\rvert})\in\Omega^fS^*$, we have $i(f)(x_1,\ldots,x_{\lvert f\rvert})\in S^{\alpha+1}$.
   Thus $S^*$ is closed for each operation of $\Sigma_0$.
   Similarly, we could have $S^*$ is closed for the operations $i(g)$ for each $g\in\Sigma_1$.
\qed

With the idea of generating the sub o.b.d.algebra closure in the proof of Proposition \ref{GenSC}, we have the following lemma by induction.

\begin{lem}
\label{RstrctandCoRstrct}
  Let $f:(P,i)\ra (Q,j)$ be a morphism of $\mathbf{DCPO}(\mathbb{T} ,\Omega)$ and $S$ a subset of $P$.
   Then the restriction and co-restriction $f|^{\circ}_{\circ}:cl^{(\mathbb{T} ,\Omega)}_{(P,i)}(S)\ra cl^{(\mathbb{T} ,\Omega)}_{(Q,j)}(f(S))$ is well-defined and it is a morphism of $\mathbf{DCPO}(\mathbb{T} ,\Omega)$.
\end{lem}

A monad on a category $\mathbf{C}$ is a triple $(\mathcal{P},\eta,\mu)$ consisting of an endofunctor $\mathcal{P}$ on $\mathbf{C}$ and the natural transformations $\eta:\mathcal{ID}_{\mathbf{C}}\ra\mathcal{P},\mu:\mathcal{P}^2\ra\mathcal{P}$ such that $\mu_A\circ\mathcal{P}\eta_A=Id_{\mathcal{P}A}=\mu_A\circ\eta_{\mathcal{P}A}$ and $\mu_A\circ\mathcal{P}\mu_A=\mu_A\circ\mu_{\mathcal{P}A}$ for every object $A$ of $\mathbf{C}$.
 Then, there is an alternative description of a monad.

A Kleisli triple on a category $\mathbf{C}$ is a triple $(\mathcal{P},\eta,\dagger)$, where $\mathcal{P}:Object(\mathbf{C})\ra Object(\mathbf{C})$, $\eta_A:A\ra\mathcal{P}A$ for $A\in Object(\mathbf{C})$, $f^\dagger:\mathcal{P}A\ra\mathcal{P}B$ for $f:A\ra\mathcal{P}B$ and the following equations hold:
  \begin{enumerate}[(i)]
    \item $\eta^{\dagger}_A=id_A$,
    \item $f^{\dagger}\circ\eta_A=f$,
    \item $f^{\dagger}\circ g^{\dagger}=(f^{\dagger}\circ g)^{\dagger}$.
  \end{enumerate}

Every Kleisli triple $(\mathcal{P},\eta,\dag)$ corresponds to a monad $(\mathcal{P},\eta,\mu)$ by defining $\mathcal{P}(f:A\ra B)=(\eta_B\circ f)^{\dagger}$ and $\mu_A=id_{\mathcal{P}A}^{\dagger}$.

\begin{defi}
\label{StrgMonad}
\cite{Moggi1991}
  Let $(\mathcal{P},\eta,\dagger)$ be a Kleisli triple on a category $\mathbf{C}$.
   If $\mathbf{C}$ has finite products and there is a natural transformation $t_{A,B}:A\times\mathcal{P}B\ra\mathcal{P}(A\times B)$ such that
  \begin{enumerate}[(i)]
    \item $\mathcal{P}(r_A)\circ t_{1,A}=r_{\mathcal{P}A}$,
    \item $\mathcal{P}(a_{A,B,C})\circ t_{A\times B,C}=t_{A,B\times C}\circ (id_A\times t_{B,C})\circ a_{A,B,\mathcal{P}C}$,
    \item $t_{A,B}\circ(id_A\times\eta_B)=\eta_{A\times B}$,
    \item $t_{A,B}\circ (id_A\times\mu_B)=\mu_{A\times B}\circ \mathcal{P}(t_{A,B})\circ t_{A,\mathcal{P}B}$,
  \end{enumerate}
  then we say that the Kleisli triple (resp., the monad) is strong, where $r_A:1\times A\ra A$ and $a_{A,B,C}:(A\times B)\times C\ra A\times (B\times C)$ are isomorphisms.
\end{defi}

In \cite{Moggi1991}, Moggi gave a convenient approach to judging whether a monad on a category is strong, we apply it to the category of dcpos, that is, the following lemma is true.

\begin{lem}
\label{EquaOfStrong}
  Suppose $(\mathcal{P},\eta,\mu)$ is a monad on $\mathbf{DCPO}$ and $A,B$ are dcpos.
   Then the monad is strong, and the strength is given by
   \begin{center}
     $t_{A,B}(a,u)=\mathcal{P}(\lambda b.(a,b))(u)$,
   \end{center}
   where $\lambda b.(a,b):B\ra A\times B$ maps every $b\in B$ to $(a,b)\in A\times B$.
\end{lem}

Next we introduce what is a monad (resp., a Kleisli triple) having algebraic features.
\begin{defi}
  Let $\mathbf{D}$ be a subcategory of a category $\mathbf{C}$, $U:\mathbf{D}\ra\mathbf{C}$ the forgetful functor and $(\mathcal{P},\eta,\dagger)$ a Kleisli triple on $\mathbf{C}$.
  Then we say that $(\mathcal{P},\eta,\dagger)$ arrives at $\mathbf{D}$ by an approach $R$ iff $R$ is an approach which could identify $\mathcal{P}$ as a functor from $\mathbf{C}$ to $\mathbf{D}$, that is:
  \begin{enumerate}[(i)]
    \item $R\mathcal{P}A$ is an object of $\mathbf{D}$ for each $A\in \mathbf{C}$, and $UR\mathcal{P}A=\mathcal{P}A$,
    \item if $A,B$ are in $\mathbf{C}$ and $f:A\ra B$ is in $\mathbf{C}$, then $Rf^{\dagger}\in\mathbf{D}$ and $URf^{\dagger}=f^{\dagger}$.
  \end{enumerate}
  Moreover, if objects and morphisms of the subcategory $\mathbf{D}$ are algebras and homomorphisms, then we say that $(\mathcal{P},\eta,\dagger)$ has algebraic features.
\end{defi}

\begin{thm}
  Let $(\mathcal{P},\eta,\dagger)$ be a Kleisli triple on $\mathbf{DCPO}$ that arrives at $\mathbf{DCPO}\mathbb{T}$ by an approach $R$, and let
  \begin{enumerate}[(i)]
    \item $\mathcal{P}^*L=cl^{(\mathbb{T} ,\Omega)}_{R\mathcal{P}A}(\eta(L))$,
    \item $\eta*_L$ be co-restriction of $\eta_L$ on $\mathcal{P}^*L$,
    \item $f^{\dagger *}:\mathcal{P}^*L\ra \mathcal{P}^*M$ is the restriction and co-restriction of $(inc^*_L\circ f)^\dagger$, where $f:L\ra \mathcal{P}^*M$ and $inc^*_L:\mathcal{P}^*L\ra\mathcal{P}L$ is the inclusion map,
    \item the interpretation of $R^*\mathcal{P}^*L$ is induced by that of $R\mathcal{P}L$, and $R^*f^{\dagger *}=f^{\dagger *}$.
  \end{enumerate}
  Then $(\mathcal{P}^*,\eta^*,\dagger*)$ is a Kleisli triple that arrives at $\mathbf{DCPO}(\mathbb{T} ,\Omega)$ by the approach $R^*$.\par
  Moreover, if $(\mathcal{P},\eta,\dagger)$ is strong, then $\mathcal{P}^*$ is strong by what the strength $t_{A,B}^*: A\times \mathcal{P}^*(B)\ra\mathcal{P}^*(A\times B)$ exactly is the restriction and co-restriction of $t_{A,B}$.
\end{thm}
\proof
  Suppose $f:L\ra\mathcal{P}^*M$ and $g:M\ra\mathcal{P}^*N$ are Scott continuous maps.
   We list the following calculations to show that $(\mathcal{P}^*,\eta^*,\dagger*)$ is a well-defined Kleisli triple.
  \begin{align}
   f^{\dagger*}\circ\eta^*_L=& (inc_M^*\circ f)^\dagger|^\circ_\circ\circ\eta_L|^\circ& \hspace{1em} (\eta_L^*)^{\dagger*}=& (inc_L^*\circ\eta_L^*)^\dagger|^\circ_\circ \nonumber\\
                            =& ((inc_M^*\circ f)^\dagger\circ\eta_L)|^\circ,         &                                   =& (\eta_L)^\dag|^\circ_\circ    \nonumber\\
                            =& (inc_M^*\circ f)|^\circ                            &                                   =& id_{\mathcal{P}L}|^\circ_\circ\nonumber\\
                            =&  f,                                            &                                   =& id_{\mathcal{P}^*L},\nonumber
  \end{align}
  \begin{align}
    g^{\dagger*}\circ f^{\dagger*}= & (inc_N^*\circ g)^\dagger|^\circ_\circ\circ(inc_M^*\circ f)^\dagger|^\circ_\circ  \nonumber\\
    = & ((inc_N^*\circ g)^\dagger\circ(inc_M^*\circ f)^\dagger)|^\circ_\circ             \nonumber\\
    = & ((inc_N^*\circ g)^\dagger\circ inc_M^*\circ f)^\dagger|^\circ_\circ              \nonumber\\
    = & ((inc_N^*\circ g)^\dagger|_\circ\circ f)^\dagger|^\circ                          \nonumber\\
    = & (inc_N^*\circ(inc_N^*\circ g)|^\circ_\circ\circ f)^\dagger|^\circ_\circ          \nonumber\\
    = & ((inc_N^*\circ g)|^\circ_\circ\circ f)^{\dagger*}                                \nonumber\\
    = & (g^{\dagger*}\circ f)^{\dagger*}.                                                 \nonumber
  \end{align}
  Then $R^*\mathcal{P}^*L$ is in $\mathbf{DCPO}(\mathbb{T},\Omega)$ by Remark \ref{SubObdaIsObda}.

  Let $\lambda b.(a,b)$ be the given map of Lemma \ref{EquaOfStrong} for some dcpo $A$ and $B$.
   We obtain $t_{A,B}(A\times\eta_B(B))=\eta_{A\times B}(A\times B)$ by (\romannumeral4) of Definition \ref{StrgMonad}.
   Since $Rt_{A,B}(a,-)=R\mathcal{P}(\lambda b.(a,b))$ is a morphism of $\mathbf{DCPO}\mathbb{T}$ for every fixed $a\in A$, we have $t_{A,B}(A,\mathcal{P}^*B)\subseteq \mathcal{P}^*(A\times B)$.
   Then, the image of $t_{A,B}$ restricted on $A\times \mathcal{P}^*B$ is included in $\mathcal{P}^*(A\times B)$.
   Hence,
  \begin{align}
    \mathcal{P}^*(\lambda b.(a,b)) = & (\eta_{A\times B}^*\circ\lambda b.(a,b))^{\dagger*} \nonumber\\
                                             = & (inc_{A\times B}^*\circ\eta_{A\times B}^*\circ\lambda b.(a,b))^\dagger|^\circ_\circ \nonumber\\
                                             = & (\eta_{A\times B}\circ\lambda b.(a,b))^\dagger|^\circ_\circ \nonumber\\
                                             = & \mathcal{P}(\lambda b.(a,b))|^\circ_\circ \nonumber\\
                                             = & t_{A\times B}(a,-)|^\circ_\circ \nonumber\\
                                             = & t_{A\times B}^*(a,-). \nonumber
  \end{align}

\qed

In the next several sections, we simplify each dcpo $\mathbb{T}$-algebra $(L,i)$ as $L$ with some given operations for each $D$-algebraic theory $\mathbb{T}$.

\section{Consistent Plotkin index power domain}
\label{SecIndex}

For convenience, we use $(a_i)$ to represent a string $(a_1,a_2,\ldots,a_n)$ with the length $n$ omitted.
 If it is necessary to emphasize the length $n$, we write $(a_i)_n$ or $(a_1,a_2,\ldots,a_n)_n$ for it.

The notion of the index power domains was introduced by Varacca in his Ph.d dissertation (\cite{Varacca2003}), and Mislove revisited it in \cite{Mislove2007}.
 Let $a$ and $a'$ be the strings of a poset $P$. Define $a\equiv_n a'$ iff both $a$ and $a'$ have length $n$ and there is a permutation from $a$ to $a'$.
 A simple example is that $(a_1,a_2,a_3)\equiv_3 (a_2,a_3,a_1)$, where $\{a_1,a_2,a_3\}\subseteq P$.
 Given a string $a=(a_i)_n$, then we write $p_i(a)$ for $a_i$, where $i=1,\ldots,n$.
 The relation $\equiv_n$ is an equivalent relation for each natural number $n\in\mathbb{N}$.

Let $Q$ be a continuous poset.
 We let $B_+(Q)=\bigcup\{P^n/\equiv_n:0<n<\omega\}\cup\{\bot\}$, and we define a relation $\ll_+$ on $B_+(Q)$ by
 \begin{enumerate}[(i)]
   \item $[(x_i)_n]\ll_+[(y_j)_m]$ iff there is an injective map $\psi:\{1,2,\ldots,n\}\ra\{1,2,\ldots,m\}$ such that $x_i\ll y_{\psi(i)}$ for $i=1,2,\ldots,n$.
   \item $\bot\ll_+[(x_i)_n]$ for every $[(x_i)_n]$, and $\bot\ll_+\bot$.
 \end{enumerate}
 Then $(B_+(Q),\ll_+)$ is an abstract basis.

Set $L$ a domain.
 The Plotkin index power domain over $L$ is denoted as $IV_p(L)$, and is given by the round ideal completion $RId(B_+(\mathbb{R}_+\times L),\ll_+)$, where $\mathbb{R}_+$ is the set of positive real numbers with the usual order.
 It is noted that $(r,x)\ll(s,y)$ in continuous poset $\mathbb{R}_+\times L$ iff $r<s$ and $x\ll y$.
 For each subset $S$ of $B_+(\mathbb{R}_+\times L)$, we write $\Da S$ for the $\ll_+$-lower set $\{[(r_i,x_i)]:\exists[(s_j,y_j)]\in S\text{ s.t. }[(r_i,x_i)]\ll_+[(s_j,y_j)]\}$ and we abbreviate $\Da\{[(r_i,x_i)]\}$ as $\Da [(r_i,x_i)]$.

The $\mathbb{QC}$ (dcpo quasi-cone) is an $D$-algebraic theory which components are same as the dcpo cone theory, except for the lack of the law $r\star(x+ y)=r\star x+ r\star y$.
 Then, the Plotkin index power domain $IV_p(L)$ over each domain $L$ gives a free construction $IV_p$ from $\mathbf{DOM}$ to $\mathbf{DCPO}\mathbb{QC} $, and the unit $\eta_L:L\ra IV_p(L)$ maps $x$ to $\Da[(1,x)]$.

Generally, each non-bottom element of $B_+(\mathbb{R}_+\times L)$ is recognized not only as an equivalent class of strings of $\mathbb{R}_+\times L$, but also as an equivalent class of $\mathbb{R}_{\geq 0}\times L$ by ignoring some pairs with index 0.
 For example, $[(1,x)]$ could be regarded as $[((0,y),(1,x))]$, $[((0,z),(0,y),(1,x))]$, etc.
 The bottom of $B_+(\mathbb{R}_+\times L)$ could be seen as the empty string or a string of $0\times L$ in arbitrary length.
 Then, operations in $IV_p(L)$ are defined by
 \begin{enumerate}[(i)]
   \item $R_1+R_2=\Da\{[(r_i,x_i)\diamond(y_j,s_j)]:[(r_i,x_i)]\in R_1,[(y_j,s_j)]\in R_2\}$, where $\diamond$ is the symbol of concatenations of strings,
   \item for each $R\in IV_p(L)$ and each $r\in\mathbb{R}_{\geq 0}$, $rR$ (we abbreviate $r\star R$ to $rR$) is given by $\Da\{[(rr_i,x_i)]:[(r_i,x_i)]\in R\}$, and
   \begin{center}
     $\infty R=\bigsqcup\{rR:r\in\mathbb{R}_{\geq 0}\}=\bigcup\{rR:r\in\mathbb{R}_{\geq 0}\}=\Da\{[(rr_i,x_i)]:[(r_i,x_i)]\in R,r\in\mathbb{R}_{\geq 0}\}$,
   \end{center}
   \item $\underline{0}=\{\bot\}$.
 \end{enumerate}

\begin{rem}
  Let $L$ be a domain and $R$ a round ideal of $IV_p(L)$.
   Then, for each $r\in\mathbb{R}_{\geq 0}$, $\{[(rr_i,x_i)]:[(r_i,x_i)]\in R\}$ is co-final in $rR$, and $\{[(r'r_i,x_i)]:[(r_i,x_i)]\in R,r'\in\mathbb{R}_{\geq 0}\}$ is co-final in $\infty R$.
   Similarly, for $R_1,R_2\in IV_p(L)$, $\{[(r_i,x_i)\diamond(y_j,s_j)]:[(r_i,x_i)]\in R_1,[(y_j,s_j)]\in R_2\}$ is co-final in $R_1+R_2$.
\end{rem}

\begin{rems}
\label{IndexOpRemark}
  Let $L$ be a domain and $[(r_i,x_i)_m],[(s_j,y_j)_n]$ arbitrary elements of $B_+(\mathbb{R}_+\times L)$.
   Then the following statements hold:
  \begin{enumerate}[(i)]
    \item if there is an injective map $\psi:\{1,2,\ldots,m\}\ra\{1,2,\ldots,n\}$ such that $r_i\leq s_{\psi(i)}$ and $x_i\leq y_{\psi(i)}$ for $i=1,2,\ldots,m$, then $\Da[(r_i,x_i)_m]\subseteq\Da[(s_j,y_j)_n]$.
    \item $\Da[(r_i,x_i)]+\Da[(s_j,y_j)]=\Da[(r_i,x_i)\diamond(s_j,y_j)]$,
    \item $\Da[(rr_i,x_i)]=r\Da[(r_i,x_i)]$ for each $r\in\mathbb{R}_{\geq 0}$,
    \item $\{[(r_i',x_i')_n]:r_i'<r_i,x_i'\ll x_i,i=1,2,\ldots,n\}$ is co-final in $\Da[(r_i,x_i)_n]$.
  \end{enumerate}
\end{rems}

\begin{prop}
\label{EmbeddingOfEta}
  Suppose $L$ is a domain.
   For each $\Da[(r_i,x_i)_m]$ and each $\Da[(s_j,y_j)_n]$ of $IV_p(L)$, if the relation $\Da[(r_i,x_i)_m]\subseteq\Da[(s_j,y_j)_n]$ holds, then there is an injective map
  \begin{center}
    $\psi:\{1,2,\ldots,m\}\ra\{1,2,\ldots,n\}$
  \end{center}
  such that $r_i\leq s_{\psi(i)}$ and $x_i\leq y_{\psi(i)}$ for $i=1,2,\ldots,m$.
\end{prop}
\proof
  It is easy to have $\{[(r_i',x_i')_m]:r_i'<r_i,x_i'\ll x_i,i=1,2,\ldots,m\}\subseteq\Da[(r_i,x_i)_m]\subseteq\Da[(s_j,y_j)_n]$.
   It implies that the subset $\{(r,x)\in\overline{\mathbb{R}}_{\geq 0}\times L:\exists (r_{i_0},x_{i_0}), r<r_{i_0},x\ll x_{i_0}\}$ is included in the Scott closed subset $\da\{(s_1,y_1),(s_2,y_2),\ldots,(s_n,y_n)\}$ of $\overline{\mathbb{R}}_{\geq 0}\times L$.
   Thus, for each $i_0$ of $\{1,2,\ldots,m\}$, the directed subset $\{(r_{i_0}',x_{i_0}'):r_{i_0}'< r_{i_0},x_{i_0}'\ll x_{i_0}\}$ of $\overline{\mathbb{R}}_{\geq 0}\times L$ is contained in the Scott closed subset $\da\{(s_1,y_1),(s_2,y_2),\ldots,(s_n,y_n)\}$.
   Hence, we have $(r_{i_0},x_{i_0})$ is in $\da\{(s_1,y_1),(s_2,y_2),\ldots,(s_n,y_n)\}$ since $\{(r_{i_0}',x_{i_0}'):r_{i_0}'<r_{i_0},x_{i_0}'\ll x_{i_0}\}$ has a sup $(r_{i_0},x_{i_0})$.
   It follows that each $(r_i,x_i)$ is lower than some $(s_{j_i},y_{j_i})$ of $[(s_j,y_j)_n]$.
   Then, we let $\psi$ be a map with $\psi(i)=j_i$, which satisfies the condition as desired.
\qed

\begin{defi}
\cite{Zou2018}
  We say that a dcpo $L$ has one-step property iff for each subset $S\subseteq L$, the Scott closure $\bar{S}$ is equal to $\mathcal{D}_s(\da S)$.
\end{defi}

\begin{prop}
\cite{Zou2018}
  Every domain has the one-step property.
\end{prop}

\begin{lem}
\cite{Goubault2013}
\label{ExtMapFromBas}
  Let $L$ be a domain, $B$ a basis of $L$ and $f$ an order-preserving map from $B$ to a dcpo $M$. Then $f$ could be extended to the greatest Scott continuous map $\tau(f)$ from $L$ to $M$ such that $\tau(f)(b)\leq f(b)$ for all $b\in B$. The extending map $\tau(f)$ is given by mapping $x$ to $\bigsqcup f(\dda x\cap B)$.
\end{lem}

\begin{defi}
   Let $\Omega^+(P)=\{(x,y):\{x,y\}^u\neq\emptyset\}$ and $\Omega=\{\Omega^+\}$.
    We call each object of $\mathbf{DCPO}(\mathbb{QC} ,\Omega)$ a consistent dcpo quasi-cone.
\end{defi}

\begin{rem}
\label{ConstCondTransToBasis}
  For every domain $D$ with a basis $B$, the consistent conditions could be transferred to the basis, that is, if $x,y\leq z$ in $D$, then for every $b_x,b_y$ of $\dda x\cap B,\dda y\cap B$, $b_x,b_y$ has an upper bound $b_z\in \dda z\cap B$.

  Similarly, the consistent condition could be transferred from round ideals to the abstract basis. For every round ideal $R_1,R_2$ of the round ideal completion $RId(A,\prec)$, if there is a $R_3$ of $RId(A,\prec)$ containing $R_1,R_2$, then every $x$ of $R_1$ and $y$ of $R_2$ is $\prec$-lower than some $z$ of $R_3$.
\end{rem}

\begin{thm}
  Let $L$ be a domain and $B$ the subset
  \begin{center}
    $\{(r_i,x_i)_n\in B_+(L):n\in\mathbb{N},\{x_1,x_2,\ldots,x_n\}^u\neq\emptyset\}$.
  \end{center}
   Then the consistent Plotkin index power over $L$, denoted as $IV^c_p(L)$, is the Scott closure of $\mathcal{B}$ carried induced order, where $\mathcal{B}=\{\Da[(r_i,x_i)]:[(r_i,x_i)]\in B\}\cup\{\{\bot\}\}$.
   The unit $\eta^c_L:L\ra IV^c_p(L)$ is the co-restriction of $\eta_L$, and $IV^c_p(L)$ over each domain $L$ gives a free construction $IV^c_p$ from $\mathbf{DOM}$ to the category of the consistent dcpo quasi-cones $\mathbf{DCPO}(\mathbb{QC} ,\Omega)$.
\end{thm}
\proof
  Notice that $IV_p(L)$ is a domain and each Scott closed subset of a domain is a domain, we have $IV^c_p(L)$ is a domain.

  \textbf{Claim 1}: $IV^c_p(L)=\{R\in IV_p(L):\forall [(r_i,x_i)_n]\in R, [(r_i,x_i)_n]\in B\}$.

  Assume $R$ is on the right.
   Then $R$ is a sup of $\{\Da[(r_i,x_i)]:[(r_i,x_i)]\in R\}$ by (\romannumeral3) of Proposition \ref{AbsBasisPPT}.
   It follows that $R\in IV^c_p(L)$ from $\{\Da[(r_i,x_i)]:[(r_i,x_i)]\in R\}\subseteq \mathcal{B}\subseteq IV^c_p(L)$.

  Conversely, suppose $R$ is in $\da \mathcal{B}$.
   There must be a $[(s_j,y_j)_n]\in B$ with $R\subseteq \Da[(s_j,y_j)_n]$.
   For each $[(r_i,x_i)]$ of $R$ with $[(r_i,x_i)]\ll_+[(s_j,y_j)_n]$, we have $[(r_i,x_i)]\in B$ by the definition of $\ll_+$.
   Thus $R$ is composed of some elements of $B$, $R$ is on the right.
   Since $IV^c_p(L)=\mathcal{D}_s(\da\mathcal{B})$ by one-step property, each $R'$ of $IV^c_p(L)$ is a sup of some directed subsets of $\da B$.
   Hence, each $R'\in IV^c_p(L)$ must comprise some elements of $B$.

  With the proof of Claim 1 completed, $IV^c_p(L)=\mathcal{D}_s(B)$, and $B$ forms a basis of $IV^c_p(L)$ by (\romannumeral4) of Proposition \ref{AbsBasisPPT}.

  \textbf{Claim 2}: $IV^c_p(L)$ is closed under operations.

  Certainly, the zero $\underline{0}=\{\{\bot\}\}$ certainly is in $IV^c_p(L)$.

  Obviously, for each $[(r_i,x_i)]$ of $B$ and each $r$ of $\mathbb{R}_{\geq 0}$, $[(rr_i,x_i)]$ is in $B$.
   We show that $rR$ is in $IV^c_p(L)$ for each $r\in\overline{\mathbb{R}}_{\geq 0}$ and each $R\in IV^c_p(L)$.
   Given an arbitrary $R$ of $IV^c_p(L)$.
   If $r\in\mathbb{R}_{\geq 0}$, then $rR=\Da\{[(rr_i,x_i)]:[(r_i,x_i)\in R]\}$ must be included by $IV^c_p(L)$ from the proof of Claim 1.
   If $r=\infty$, then $\infty R=\bigsqcup \{r'R:r'\in\mathbb{R}_{\geq 0}\}$ is in $IV^c_p(L)$ since $IV^c_p(L)$ is Scott closed.

  Let $\{R_1,R_2,R_3\}$ be a subset of $IV^c_p(L)$ with $R_1,R_2\subseteq R_3$, $[(r_i,x_i)]$ an arbitrary element of $R_1$ and $[(s_j,y_j)]$ an arbitrary element of $R_2$.
   Then there is a $[(t_k,z_k)]\in R_3$ with $[(r_i,x_i)],[(s_j,y_j)]\ll_+[(t_k,z_k)]$ by Remark \ref{ConstCondTransToBasis}.
   Since $[(r_i,x_i)\diamond(s_j,y_j)]\ll_+[(t_k,z_k)]$ from the definition of $\ll_+$, we have $[(r_i,x_i)\diamond(s_j,y_j)]\in B$.
   Thus $R_1+R_2$, which is equal to
   \begin{center}
     $\Da\{[(r_i,x_i)\diamond(s_j,y_j)]:[(r_i,x_i)]\in R_1,[(s_j,y_j)]\in R_2\}$,
   \end{center}
   is composed of some elements of $B$.
   Hence, $R_1+R_2$ is in $IV^c_p(L)$.

  \textbf{Claim 3}: $IV^c_p(L)$ is exactly  $cl^{(\mathbb{T} ,\Omega)}_{IV_p(L)}(\eta_L(L))$, the sub o.b.d.algebra of $\eta_L(L)$.

  Given a $[(r_i,x_i)_n]$ of $B$ and an upper bound $x$ of $\{x_1,x_2,\ldots,x_n\}$.
   By Remark \ref{IndexOpRemark}, we can easily obtain $\Da[(r_i,x_i)]=r_i\Da[(1,x_i)]$ for $i=1,2,\ldots,n$.
   Then, $\Da[(r_i,x_i)]$ is in $cl^{(\mathbb{T} ,\Omega)}_{IV_p(L)}(\eta_L(L))$ for $i=1,2,\ldots,n$, because $cl^{(\mathbb{T} ,\Omega)}_{IV_p(L)}(\eta_L(L))$ is closed under the scalar multiplication.

  Clearly, $\Da[((\Sigma_{i=1}^nr_i,x),(\Sigma_{i=1}^nr_i,x),\ldots,(\Sigma_{i=1}^nr_i,x))_n]$ is in $cl^{(\mathbb{T} ,\Omega)}_{IV_p(L)}(\eta_L(L))$.
   Therefore,
   \begin{center}
     $\Da[(r_i,x_i)_n]=\Da[(r_1,x_1)]+\Da[(r_2,x_2)]+\ldots+\Da[(r_n,x_n)]$
   \end{center}
   is in $cl^{(\mathbb{T} ,\Omega)}_{IV_p(L)}(\eta_L(L))$, where the sum exists since $\Da[((\Sigma_{i=1}^nr_i,x),(\Sigma_{i=1}^nr_i,x),\ldots,(\Sigma_{i=1}^nr_i,x))_n]$ is an upper bound of $\{\Da[(r_i,x_i)]:i=1,2,\ldots,n\}$ from (\romannumeral1) of Remark \ref{IndexOpRemark}.
   It implies that $\mathcal{B}$ is included by $cl^{(\mathbb{T} ,\Omega)}_{IV_p(L)}(\eta_L(L))$.
   So, each $R$ of $IV^c_p(L)$ is a sup of some directed subset of $\mathcal{B}$.
   Thus, $IV^c_p(L)$ is contained in $cl^{(\mathbb{T} ,\Omega)}_{IV_p(L)}(\eta_L(L))$.
   As a sub o.b.d.algebra containing $\eta_L(L)$, $IV^c_p(L)$ is exactly $cl^{(\mathbb{T} ,\Omega)}_{IV_p(L)}(\eta_L(L))$.

  Given a Scott continuous map $f$ from $L$ to a consistent dcpo quasi-cone $Q$.
   For each $[(r_i,x_i)_n]\in B$ and each upper bound $x$ of $\{x_1,x_2,\ldots,x_n\}$, $\Sigma_{i=1}^nr_if(x)$ is an upper bound of the set of $r_1f(x_1),r_2f(x_2),\ldots,r_nf(x_n)$.
   Thus $\Sigma_{i=1}^nr_if(x_i)$ exist in $Q$.

  We define a map $\hat{f}_0$ from the basis $\mathcal{B}$ to $Q$ by $\hat{f}_0(\Da[(r_i,x_i)])=\Sigma_{i=1}^nr_if(x_i)$.
   By Proposition \ref{EmbeddingOfEta}, if there is $\Da[(r_i,x_i)_m]\subseteq\Da[(s_j,y_j)_n]$, then it is easy to have $\Sigma_{i=1}^mr_if(x_i)\leq\Sigma_{j=1}^ns_jf(y_j)$.
   Thus $\hat{f}_0$ preserves the order.
   We let $\hat{f}:IV^c_p(L)\ra Q$ be the extending map of $\hat{f}_0$ from Lemma \ref{ExtMapFromBas}.

  Select a round ideal $R$ of $IV^c_p(L)$.
  By Proposition \ref{AbsBasisPPT}, $\{\Da[(r_i,x_i)]:[(r_i,x_i)]\in R\}$ is a directed subset in which each element is way-below $R$, and it has a sup $R$.
   Then, we have
   \begin{center}
     $\hat{f}(R)=\bigsqcup\{\Sigma_{i=1}^nr_if(x_i):[(r_i,x_i)_n]\in R\}$.
   \end{center}

  For each $\{R_1,R_2,R_3\}$ of $IV^c_p(L)$ with $R_1,R_2\subseteq R_3$, we calculate
  \begin{align}
    \hat{f}(R_1+R_2) & = \bigsqcup\{\hat{f}_0(\Da[(t_k,z_k)]):[(t_k,z_k)]\in R_1+R_2\} \nonumber\\
                     & = \bigsqcup\{\hat{f}_0(\Da[(r_i,x_i)\diamond(s_j,y_j)]):[(r_i,x_i)]\in R_1,[(s_j,y_j)]\in R_2\}\nonumber\\
                     & = \bigsqcup\{\Sigma_{i=1}^mr_if(x_i)+\Sigma_{j=1}^ns_jf(y_j):[(r_i,x_i)_m]\in R_1,[(s_j,y_j)_n]\in R_2 \} \nonumber\\
                     & = \bigsqcup\{\Sigma_{i=1}^mr_if(x_i):[(r_i,x_i)_m]\in R_1\}+\bigsqcup\{\Sigma_{j=1}^ns_jf(y_j):[(s_j,y_j)_n]\in R_2\}\nonumber \\
                     & = \hat{f}(R_1)+\hat{f}(R_2).\nonumber
  \end{align}

  For each $R\in IV^c_p(L)$ and $r\in\overline{\mathbb{R}}_{\geq 0}$, if $r\in\mathbb{R}_{\geq 0}$, then we have
  \begin{align}
    \hat{f}(rR) & = \bigsqcup \{\Sigma_{i=1}^mrr_if(x_i):[(r_i,x_i)_m]\in R\}\nonumber\\
                &  = r\bigsqcup \{\Sigma_{i=1}^mr_if(x_i):[(r_i,x_i)_m]\in R\}\nonumber\\
                &  = r\hat{f}(R),\nonumber
  \end{align}
  and if $r=\infty$, then
  \begin{align}
    \hat{f}(\infty R) & = \bigsqcup \{\Sigma_{i=1}^mr'r_if(x_i):[(r_i,x_i)_m]\in R,r'\in\mathbb{R}_{\geq 0}\} \nonumber\\
                      & = \bigsqcup_{r'\in\mathbb{R}_{\geq 0}}\bigsqcup\{\Sigma_{i=1}^mr'r_if(x_i):[(r_i,x_i)_m]\in R,r'\in\mathbb{R}_{\geq 0}\} \nonumber\\
                      & = \bigsqcup_{r'\in\mathbb{R}_{\geq 0}}\hat{f}(r'R)\nonumber \\
                      & = \bigsqcup_{r'\in\mathbb{R}_{\geq 0}} r'\hat{f}(R)\nonumber \\
                      & = \infty \hat{f}(R).\nonumber
  \end{align}

  By  (\romannumeral4) of Remark \ref{IndexOpRemark}, $\{\Sigma_{i=1}^nr_i'f(x_i'):r_i'<r_i,x_i'\ll x_i,i=1,2,\ldots,n\}$ is co-final in $\{\Sigma_{i=1}^ms_jf(y_j):[(s_j,y_j)_m]\ll_+[(r_i,x_i)_n]\}$ for each $[(r_i,x_i)_n]\in B$.
   It follows that
   \begin{center}
     $\hat{\Da[(r_i,x_i)_n]}=\bigsqcup\{\Sigma_{i=1}^nr_i'f(x_i'):r_i'<r_i,x_i'\ll x_i,i=1,2,\ldots,n\}=\Sigma_{i=1}^nr_if(x_i)$
   \end{center}
   by continuity of operations in $Q$.
   Thus, $\hat{f}\circ\eta^c_L=f$.

  It remains to show the uniqueness of $\hat{f}$.
   Suppose that there is another homomorphism $g:IV^c_p(L)\ra Q$ with $g\circ\eta_L^c=f$.
   Then, $g(\Da[(1,x)])=f(x)$ for every $x\in L$.
   So, $g(\Da[(r_i,x_i)]_n)=\Sigma_{i=1}^n r_if(x_i)$ for every $[(r_i,x_i)]_n\in B$ since $g$ preserves operations.
   Furthermore, $g(R)=\bigsqcup\{\Sigma_{i=1}^n r_if(x_i):[(r_i,x_i)]_n\}$ by the continuity of $g$.
   We conclude that $g=f$.
\qed

\section{Consistent probabilistic power domain}
\label{SecKeg}
Kegelspitzen was introduced by Plotkin and Keimel for describing algebraic features of the probabilistic power construction (\cite{Klaus2017}).
 The kegelspitze theory $\mathbb{K}=(\Sigma,\Sigma_1,\lvert\cdot\rvert,M,\theta,\Lambda)$ is an $D$-algebraic theory, where
  \begin{enumerate}[(i)]
    \item $\Sigma_0=\{\underline{0},+_r:r\in[0,1]\}$,
    \item $\Sigma_1=\{\star\}$,
    \item $\lvert\underline{0}\rvert=0$, $\lvert\star\rvert=2$, and $\lvert+_r\rvert=2$ for each $r\in[0,1]$
    \item $M=\{[0,1]\}$, where $[0,1]$ endowed with the usual order,
    \item $\theta(\star,1)=[0,1]$ and $\theta(\star,2)=*$,
    \item $\Lambda$ consists of the following laws:
   \begin{align}
   &(1)\quad x +_1 y  = x                                          &          &(6)\quad 0 \star x  = \underline{0} = r \star \underline{0}  \nonumber\\
   &(2)\quad x +_r x  = x                                          &          &(7)\quad 1 \star x  = x                                      \nonumber\\
   &(3)\quad x +_r y  = y +_{1-r} x                                &            &(8)\quad (rs) \star x  = r\star (s\star x)                   \nonumber\\
   &(4)\quad(x +_r y) +_s z  = x +_{rs} (y +_{\frac{r-rs}{1-rs}} z)&            &(9)\quad r \star (x +_s y)  = r \star x +_s r \star y .        \nonumber\\
   &(5)\quad r \star x  = x +_r \underline{0}                       \nonumber
   \end{align}
  \end{enumerate}

A valuation $\mu$ on a dcpo $L$ is a map from the lattice of Scott open subsets of $L$ to $\overline{\mathbb{R}}_{\geq 0}$ and satisfies the following conditions:

 \begin{enumerate}[(i)]
   \item $\mu(\emptyset)=0$,
   \item if $U\subseteq V$ then $\mu(U)\leq\mu(V)$,
   \item $\mu(U)+\mu(V)=\mu(U\cup V)+\mu(U\cap V)$.
 \end{enumerate}

A valuation is said to be continuous iff it is a Scott continuous map.

The addition, the scalar multiplication and the zero-element of valuations are, respectively, defined by:
 \begin{enumerate}[(i)]
   \item $\mu_1+\mu_2(U)=\mu_1(U)+\mu_2(U)$,
   \item $(r\star\mu)(U)=r\mu(U)$,
   \item $\underline{0}$ is the valuation $\mu_0:\mu_0(U)=0$ for all open subset $U$.
 \end{enumerate}

For a domain $L$, the Dirac valuation $\delta_x$ of $x\in L$ maps open set $U$ to 1 if $x\in U$, or maps $U$ to 0 if $x\notin U$.
 Simple valuations are finitely linear sums of Dirac valuations, and they are continuous valuations.
 The set of continuous valuations with the pointwise order (that is, $\mu\leq\upsilon$ iff $\mu(U)\leq\upsilon(U)$ for all open set $U$) is denoted as $\mathcal{V}(L)$, and the set of continuous valuations whose images are less than 1 with the pointwise order, which is named the probabilistic power domain over $L$, is denoted as $\mathcal{V}_{\leq 1}(L)$.

The probabilistic power domain $\mathcal{V}_{\leq 1}(L)$ over each domain $L$ gives a free construction $\mathcal{V}_{\leq 1}$ from $\mathbf{DOM}$ to $\mathbf{DCPO}\mathbb{K}$.
 The unit $\eta_L:L\ra \mathcal{V}_{\leq 1}(L)$ maps $x$ to $\delta_x$.
 The operation $+_r$ in $\mathcal{V}_{\leq 1}(L)$ is defined by $\mu_1+_r\mu_2=r\mu_1+(1-r)\mu_2$.
 The scalar multiplication and zero-element are induced by the operations of valuations.
 Simple valuations form a basis for $\mathcal{V}_{\leq 1}(L)$.
 A simple valuation $\Sigma_{i\in I}r_i\delta_{x_i}$ is way-below $\mu$ in $\mathcal{V}_{\leq 1}(L)$ iff for each $S\subseteq I$, there is $\Sigma_{i\in S}r_i<\mu(\bigcup_{i\in S}\dua x_i)$.
 It is easy to deduce that the addition and the scalar multiplication keep the way-below relation.

\begin{rem}
\label{SimpValuaApprox}
  For a domain $L$, simple valuations $\Sigma_{i\in I}r_i'\delta_{x_i'}\ll \Sigma_{i\in I}r_i\delta_{x_i}$ if $r_i'<r_i$ and $x_i'\ll x_i$ for all $i\in I$.
\end{rem}

In \cite{Jones1990}, Jones constructs finitely linear sum $\Sigma_{i=1}^n r_ix_i$ (here, we abbreviate the scalar multiplication $r\star x$ to $rx$) in each kegelspitze $K$ by inductions for all $\{x_i:i=1,2,\ldots,n\}$ of $K$ and all $\Sigma_{i=1}^n r_i\leq 1$ .
 For the case that $\Sigma_{i=1}^n r_i=1$, $\Sigma_{i=1}^nr_ix_i$ is defined by $x_1+_{r_1}(\Sigma_{i=2}^n\frac{r_i}{1-r_1}x_i)$.
 In the case that $\Sigma_{i=1}^n r_i<1$, $\Sigma_{i=1}^nr_ix_i$ is defined by $\Sigma_{i=1}^n\frac{r_i}{\Sigma_{i=1}^nr_i} x_i+_{\Sigma_{i=1}^nr_i}\underline{0}$.

\begin{prop}
\label{KegelEqua}
\cite{Jones1990}
  The following equations are held in each kegelspitze by laws:
 \begin{enumerate}[(i)]
    \item $(\Sigma_{i=1}^nr_ix_i)+_t(\Sigma_{j=1}^ms_jy_j)=\Sigma_{i=1}^ntr_ix_i+\Sigma_{j=1}^m(1-t)s_jy_j$,
    \item $\Sigma_{i=1}^nr_i(\Sigma_{j=1}^{n_i}r_j^ix_j^i)=\Sigma_{i,j}r_ir_j^ix_j^i$,
    \item $\Sigma_{i=1}^nr_ix_i=\Sigma_{i=1}^nr_{\pi(i)}x_{\pi(i)}$ where $\pi$ is a permutation,
    \item $s\Sigma_{i=1}^nr_ix_i=\Sigma_{i=1}^nsr_ix_i$ for some $s\in[0,1]$.
  \end{enumerate}
\end{prop}

\begin{defi}
  Let $\Omega^{+_r}(P)=\{(x,y):\{x,y\}^u\neq\emptyset\}$, $\Omega=\{\Omega^{+_r}:r\in[0,1]\}$.
 We call each object $K$ of $\mathbf{DCPO}(\mathbb{K},\Omega)$ a consistent kegelspitze; moreover, if the carrier of $K$ is a domain, we call it a continuous consistent kegelspitze.
\end{defi}

Then we use Jones's approach to constructing $\Sigma_{i=1}^nr_ix_i$ in each $K\in\mathbf{DCPO}(\mathbb{K},\Omega)$ whenever there are $y_1,y_2,\ldots,y_n$ such that $y_i$ is greater than both $x_i$ and $\Sigma_{j=i+1}^n\frac{r_j}{1-r_1-r_2-\cdots-r_i}x_j$ for each $i\in\{1,2,\ldots,n-1\}$.
 Since every consistent kegelspitze $K$ satisfies skipped obeying the laws (see Definition \ref{Obda}), each equation composed from some of that of Proposition \ref{KegelEqua} holds in $K$ whenever both sides of the equation exist.

\begin{defi}
\label{LinearSumInConK}
  In a consistent kegelspitze $K$, we define the finitely linear sum $\Sigma_{i\in I}r_ix_i$ as an element which is equal to
  \begin{enumerate}[(i)]
    \item $\Sigma_{j=1}^{n}r_{i_j}x_{i_j}$, whenever $\Sigma_{j=1}^{n}r_{i_j}x_{i_j}$ exists and $I=\{i_1,i_2,\ldots,i_n\}$, or,
    \item $\Sigma_{i\in S}\frac{r_i}{r}x_i+_r\Sigma_{i\in I\setminus S}\frac{r_i}{1-r} r_ix_i$, whenever $S\subseteq I$, $r\in[0,1]$, both $\Sigma_{i\in S}\frac{r_i}{r}x_i$ and $\Sigma_{i\in I\setminus S}\frac{r_i}{1-r} r_ix_i$ exists and there is a $y\in K$ greater than them, or,
    \item $r\Sigma_{i\in I}\frac{r_i}{r}x_i$, whenever $\Sigma_{i\in I}\frac{r_i}{r}x_i$ exists and $r\in[0,1]$, or,
    \item $\Sigma_{a\in A}s_a(\Sigma_{b\in B_a}s_b^a x_b^a)$, whenever $\Sigma_{a\in A}s_a(\Sigma_{b\in B_a}s_b^a x_b^a)$ exists and there is a one-to-one map $\psi$ from the co-product $\coprod_{a\in A}B_a$ to $I$ such that
        \begin{center}
          $s_as^a_b=r_{\psi(a,b)}$ and $x^a_b=x_{\psi(a,b)}$.
        \end{center}
  \end{enumerate}
\end{defi}

The following lemma describes some relations between simple valuations in $\mathcal{V}_{\leq 1}(L)$.
 \begin{lem}[Splitting Lemma]\cite{Jones1990}\label{SpltingLema1}
    For two simple valuations $\Sigma_{i\in I}r_i\delta_{x_i}$ and $\Sigma_{j\in J}s_j\delta_{x_j}$ in $\mathcal{V}_{\leq 1}(L)$, $L$ a domain, we have $\Sigma_{i\in I}r_i\delta_{x_i}\leq\Sigma_{j\in J}s_j\delta_{x_j}$ iff there are non-negative real numbers $t_{ij}$ such that for each $i\in I$, $j\in J$,
     \begin{center}
       $\Sigma_{j\in J}t_{ij}=r_i$, $\Sigma_{i\in I}t_{ij}\leq s_j$,
     \end{center}
    and $t_{ij}\neq0$ implies $x_i\leq x_j$.
 \end{lem}

In what follows, the notation $\dda\mu$ always takes all continuous valuations way-below $\mu$ in $V_{\leq 1}(L)$ for every domain $L$ and every continuous valuation $\mu\in V_{\leq 1}(L)$.

\begin{prop}
  Let $L$ be a domain, $V_s(L)$ the set of simple valuations of $V_{\leq 1}(L)$ and $V_{\leq 1}^c(L)=\mathcal{D}_s(\da\eta_L(L)\cap V_s(L))$.
   Then $V_{\leq 1}^c(L)$ is a sub o.b.d.algebra of $V_{\leq 1}(L)$ and a domain; hence, it is a continuous consistent kegelspitze.
\end{prop}
\proof
  For every $\mu\in V_{\leq 1}^c(L)$, there is a directed subset $D$ of $\da\eta_L(L)\cap V_s(L)$ with $\mu=\bigsqcup D$.
   Notice that $V_s(L)$ is a basis of $V_{\leq 1}(L)$, each simple valuation of $\dda\mu$ is lower than some of $D$.
   Moreover, each simple valuation of $\dda\mu$ is lower than $\delta_y$ for some $y\in L$, that is, $\dda\mu\cap V_s(L)\subseteq\da\eta_L(L)\cap V_s(L)$.
   Suppose $A$ is a directed subset of $V_{\leq 1}^c(L)$.
   Since $\bigsqcup A=\bigsqcup\bigcup\{\dda\mu\cap V_s(L):\mu\in A\}$, $\bigsqcup A$ is a sup of a directed subset of $\da\eta_L(L)\cap V_s(L)$.
   Thus $V_{\leq 1}^c(L)$ is a subdcpo of $V_{\leq 1}(L)$, and it is a domain from the fact that $\da\eta_L(L)\cap V_s(L)$ is a basis of it.

  Assume $\mu_1,\mu_2\leq\mu_3$ in $V_{\leq1}^c(L)$.
   Then for each $\mu_1'\in\dda\mu_1\cap V_s(L)$ and each $\mu_2'\in\dda\mu_2\cap V_s(L)$, there is a $\mu_3'\in\dda\mu_3\cap V_s(L)$ is greater than them.
   Following from the fact that $\mu_3'$ is lower than some $\delta_y$, $\mu_1'$ and $\mu_2'$ have an upper bound $\delta_y$; furthermore, $r\mu_1'+(1-r)\mu_2'$ is still lower than $\delta_y$ for arbitrary $r\in[0,1]$.
   Thus,
   \begin{center}
     $\{r\mu_1'+(1-r)\mu_2':\mu_i'\in\dda\mu_i\cap V_s(L),i=1,2\}$
   \end{center}
   is a directed subset of $\da\eta_L(L)\cap V_s(L)$ and has a sup $r\mu_1+(1-r)\mu_2$ since the operations are Scott continuous.
   Moreover, this directed subset is in $\dda(r\mu_1+(1-r)\mu_2)\cap\eta_L(L)\cap V_s(L)$ because the operations preserve the way-below relation.
   It implies that $V_{\leq 1}^c(L)$ is closed under every consistent $+_r$.
   Also, it is easy to verify that $V_{\leq 1}^c(L)$ is closed under the scalar multiplication by $r\mu=\bigsqcup r(\dda\mu\cap V_s(L))$ in $V_{\leq1}(L)$.
\qed

\begin{cor}
  The basis $\eta_L(L)\cap V_s(L)$ is closed under consistent $+_r$ and the scalar multiplication.
\end{cor}

\begin{cor}
  For every $\mu_1,\mu_2\leq\mu_3$ in $V_{\leq 1}^c$,
  \begin{align}
    & r(\dda\mu_1\cap\eta_L(L)\cap V_s(L))+(1-r)(\dda\mu_2\cap\eta_L(L)\cap V_s(L))\nonumber\\
    =& \{rv_1+(1-r)v_2:v_i\in\dda\mu_i\cap\eta_L(L)\cap V_s(L),i=1,2\}\nonumber
  \end{align}
  is a co-final subset of $\dda(r\mu_1+(1-r)\mu_2)$.
  Likewise,
  \begin{center}
    $r(\dda\mu\cap\eta_L(L)\cap V_s(L))=\{rv:v\in\dda\mu\cap\eta_L(L)\cap V_s(L)\}$
  \end{center}
  is a co-final subset of $\dda r\mu\cap\eta_L(L)\cap V_s(L)$.
\end{cor}

\begin{rem}
  Utilizing Splitting Lemma \ref{SpltingLema1}, a simple valuation $\Sigma_{i\in I}r_i\delta_{x_i}$ is lower than $\delta_y$ for some $y\in L$ iff the support of it, $\{x_i\}_{i\in I}$, has $y$ as an upper bound.
  Thus $\da\eta_L(L)\cap V_s(L)$ is exactly the set of all simple valuations $\Sigma_{i\in I}r_i\delta_{x_i}$ that has a bounded support and $\Sigma_{i\in I}r_i\leq 1$.
\end{rem}

\begin{prop}
\label{ConLinExRemark}
  Let $K$ be a consistent kegelspitze, $\{x_i\}_{i\in I}$ a finite subset of $K$ and $y$ an upper bound of $\{x_i\}_{i\in I}$.
   Then $\Sigma_{i\in I}r_ix_i$ exists whenever $\Sigma_{i\in I}r_i\leq 1$.
\end{prop}
\proof
  Assume that there is a permutation such that $\{x_i\}_{i\in I}=\{x_1,x_2,\ldots,x_n\}$.
   In case $\Sigma_{i\in I}r_i=1$,
   \begin{center}
     $x_1+_{r_1}(x_2+_{\frac{r_2}{\Sigma_{i=2}^nr_i}}(\ldots(x_{n-1}+_{\frac{r_{n-1}}{r_{n-1} + r_n}}x_n)\ldots))$
   \end{center}
   exists since there is an upper bound $y$ for the pairs in the calculation in each step, and it is what $\Sigma_{i\in I}r_ix_i$ is defined.
   Also, it is similar to verify the case $\Sigma_{i\in I}r_i<1$.
\qed

Recall the constructions in inductions for generating the sub o.b.d.algebra closure, in case that $S$ is a subset of $V_{\leq 1}(L)$ over a domain $L$, then,
\begin{align}
   &S^0=S,\nonumber\\
   &S^{\alpha}_0=\bigcup_{\beta<\alpha}S^{\beta},\nonumber\\
   &S^{\alpha}_1=\{r\mu:r\in[0,1],\mu\in S^\alpha_0\},\nonumber\\
   &S^{\alpha}_2=\{r\mu_1+(1-r)\mu_2:r\in[0,1],\{\mu_1,\mu_2\}\subseteq S^\alpha_0,\{\mu_1,\mu_2\}^u\cap S^\alpha_0\neq\emptyset\},\nonumber\\
   &S^{\alpha}=\mathcal{D}_s(S^{\alpha}_0\cup S^{\alpha}_1\cup S^{\alpha}_2),\nonumber\\
   &S^*=\bigcup_{\alpha\in ORD}S^{\alpha},\nonumber
\end{align}
and obviously, $S^n=S^{n+1}_0$ for each natural number $n$.

\begin{prop}
  $V_{\leq 1}^c(L)$ is the sub o.b.d.algebra closure of $\eta_L(L)$ in $V_{\leq 1}(L)$.
\end{prop}
\proof
  For each simple valuation $\Sigma_{i\in I}r_i\delta_{x_i}$ of $\da\eta_L(L)\cap V_s(L)$, it is easy to verify that $\Sigma_{i\in I}r_i\delta_{x_i}$ must be contained in $\eta_L(L)^n$, where $n$ is the number of $I$.
  Hence, $\da\eta_L(L)\cap V_s(L)$ is included in the sub o.b.d.algebra closure of $\eta_L(L)$; moreover, $V_{\leq 1}^c(L)$ also is.
\qed

\begin{thm}
  The construction $V_{\leq 1}^c(L)$ over every domain $L$ gives a free construction $V_{\leq 1}^c$ from $\mathbf{DOM}$ to $\mathbf{DCPO}(\mathbb{K},\Omega)$, we name it the consistent power kegelspitze. The unit $\eta_L^c$ is the co-restriction of $\eta_L$.
\end{thm}
\proof
  Assume that $f$ is a Scott continuous map from $L$ to a consistent kegelspitze $K$.
   Then, we define a map $\hat{f}_0:\da\eta_L(L)\cap V_s(L)\ra K$ by
   \begin{center}
     $\hat{f}_0(\Sigma_{i\in I}r_i\delta_{x_i})=\Sigma_{i\in I}r_if(x_i)$,
   \end{center}
   where the linear sum exists in $K$ by Remark \ref{ConLinExRemark} and $f(y)\geq f(x_i)$ for each $i\in I$.

  Suppose that $\Sigma_{i\in I}r_i f(x_i)\leq\Sigma_{j\in J}s_j f(x_j)$ and both are in $\da\eta_L(L)\cap V_s(L)$.
   By Splitting Lemma \ref{SpltingLema1}, there are non-negative real numbers $t_{ij}$ with $\Sigma_{j\in J}t_{ij}=r_i$, $\Sigma_{i\in I}t_{ij}\leq s_j$, and $t_{ij}\neq0$ such  that $x_i\leq x_j$.
   So we have the following deduction:
   \begin{align}
     \Sigma_{i\in I}r_i f(x_i)=\Sigma_{i\in I}\Sigma_{j\in J}t_{ij}f(x_i)\leq\Sigma_{i\in I}\Sigma_{j\in J}t_{ij}f(x_j)=\Sigma_{j\in J}\Sigma_{i\in I}t_{ij}f(x_j)\leq\Sigma_{j\in J}s_j f(x_j).\nonumber
   \end{align}
   Since $K$ satisfies skipped obeying the laws, we conclude that $\Sigma_{i\in I}r_i f(x_i)\leq \Sigma_{j\in J}s_j f(x_j)$.
   Thus $\hat{f}_0$ preserves the order on $\da\eta_L(L)\cap V_s(L)$.
   If the simple valuations $\Sigma_{i\in I}r_x\delta_{x_i},\Sigma_{j\in J}s_j\delta_{y_j}$ are lower than some $\mu$ of $\da\eta_L(L)\cap V_s(L)$, then these simple valuations must be lower than $\delta_z$ for some $z\in L$, and then for each $t\in [0,1]$, we have
   \begin{align}
     \hat{f}_0(\Sigma_{i\in I}r_i\delta_{x_i}+_t\Sigma_{j\in J}s_j\delta_{y_j}) & = \hat{f}_0(t\Sigma_{i\in I}r_i\delta_{x_i}+(1-t)\Sigma_{j\in J}s_j\delta_{y_j})\nonumber \\
      & = \hat{f}_0(\Sigma_{i\in I}tr_i\delta_{x_i}+\Sigma_{j\in J}(1-t)s_j\delta_{y_j})\nonumber \\
      & = \Sigma_{i\in I}tr_if(x_i)+\Sigma_{j\in J}(1-t)s_jf(y_j)\nonumber\\
      & = \Sigma_{i\in I}r_if(x_i)+_t\Sigma_{j\in J}s_jf(y_j) \nonumber \\
      & = \hat{f}_0(\Sigma_{i\in I}r_i\delta_{x_i})+_t\hat{f}_0(\Sigma_{j\in J}r_i\delta_{y_j})\nonumber
   \end{align}
   by Definition \ref{LinearSumInConK}.
   So, $\hat{f}_0$ preserves every consistent $+r$.
   Also, we have $\hat{f}_0$ preserves the scalar multiplication by the formula of $\hat{f}_0$ directly.

   Let $\hat{f}:V_{\leq 1}^c(L)\ra K$ be the extending map of $\hat{f}_0$ given in Lemma \ref{ExtMapFromBas}, that is,
   \begin{center}
     $\hat{f}(\mu)=\bigsqcup\hat{f}_0(\dda\mu\cap\da\eta_L(L)\cap V_s(L))$.
   \end{center}

   We set $B^c$ denote the basis $\da\eta_L(L)\cap V_s(L)$ in the following calculations for convenience.
   For every $r\in[0,1]$ and every $\{\mu_1,\mu_2,\mu_3\}\subseteq V_{\leq 1}^c$ satisfying $\mu_1,\mu_2\leq\mu_3$, we calculate
   \begin{align}
     \hat{f}(\mu_1+_r\mu_2) = & \bigsqcup\hat{f}_0(\dda(r\mu_1+(1-r)\mu_2)\cap B^c)                                        \nonumber\\
                            = & \bigsqcup\hat{f}_0((\dda r\mu_1\cap B^c)+(\dda (1-r)\mu_2\cap B^c))                   \nonumber\\
                            = & \bigsqcup\hat{f}_0(r(\dda\mu_1\cap B^c)+(1-r)(\dda\mu_2\cap B^c))                     \nonumber\\
                            = & \bigsqcup\hat{f}_0((\dda\mu_1\cap B^c)+_r(\dda\mu_2\cap B^c))                         \nonumber\\
                            = & \bigsqcup(\hat{f}_0(\dda \mu_1\cap B^c)+_r\hat{f}_0(\dda \mu_2\cap B^c))              \nonumber\\
                            = & \bigsqcup\hat{f}_0(\dda \mu_1\cap B^c)+_r\bigsqcup\hat{f}_0(\dda \mu_2\cap B^c)       \nonumber\\
                            = & \hat{f}(\mu_1)+_r\hat{f}(\mu_2).                                                                \nonumber
   \end{align}

   For each $r\in[0,1]$ and each $\mu\in S^\omega$, we have
   \begin{align}
     \hat{f}(r\mu)   = & \bigsqcup\hat{f}_0(\dda r\mu\cap B^c)    \nonumber\\
                     = & \bigsqcup\hat{f}_0(r(\dda(\mu\cap B^c))  \nonumber\\
                     = & \bigsqcup r(\hat{f}_0(\dda \mu\cap B^c)) \nonumber\\
                     = & r\bigsqcup\hat{f}_0(\dda \mu\cap B^c)    \nonumber\\
                     = & r\hat{f}(\mu).                                \nonumber
   \end{align}

  It is necessary to show $\hat{f}=\hat{f}_0$ whenever $\hat{f}$ is restricted on $\da\eta_L(L)\cap V_s(L)$.
   For every simple valuation $\mu=\Sigma_{i\in I}r_i\delta_{x_i}$ of $\da\eta_L(L)\cap V_s(L)$, the directed subset
   \begin{center}
     $B_\mu:=\{\Sigma_{i\in I}r_i'\delta_{x_i'}:0<r_i'<r_i,x_i'\ll x_i\}$
   \end{center}
   has it as a sup since $\eta_L$ and operations are Scott continuous and $V^c_{\leq 1}(L)$ is a subdcpo; then it is easy to check that the subset is in $\dda\mu\cap\da\eta_L(L)\cap V_s(L)$ by Remark \ref{SimpValuaApprox}.
   Thus $B_\mu$ actually is co-final in $\dda\mu\cap\eta_L(L)\cap V_s(L)$.
   Hence, $\hat{f}(\mu)$ is equal to
   \begin{align}
     \bigsqcup\hat{f}_0(B_\mu) = & \bigsqcup\Sigma_{i=1}^nr_i'f(x_i')\tag{an arbitrary permutation of $I$ here}\nonumber \\
     = & \Sigma_{i=1}^nr_i'f(x_i') \tag{from partial Scott continuity of operations in $K$}\nonumber\\
     = & \hat{f}_0(\mu).\nonumber
   \end{align}
   With this result, $\hat{f}$ maps $\delta_x$ to $f(x)$, then obviously, $\hat{f}\circ\eta^c_L=f$.

  The last point is the uniqueness of $\hat{f}$.
   Suppose $g$ is another homomorphism with $g\circ\eta^c_L=f$.
   Then it has to map each $\delta_x$ to $f(x)$.
   So, $g(\Sigma_{i\in I}r_i\delta_{x_i})=\Sigma_{i\in I}r_if(x_i)$ for each $\Sigma_{i\in I}r_i\delta_{x_i}$ of $\da\eta_L(L)\cap V_s(L)$.
   Furthermore, $g(\mu)=\bigsqcup g(\dda\mu\cap \da\eta_L(L)\cap V_s(L))$ by the Scott continuity of $g$.
   We conclude that $g=f$.
\qed

\begin{exa}
  For the set $\{a,b\}$ with the discrete order,  we put the diagram of $V^c_{\leq 1}(\{a,b\})$ below.
\begin{figure}[h]
  \centering
  \begin{tikzpicture}
    \node (a)at(0,-1){$\underline{0}$};
    \node (b)at(1,1){$\delta_a$};
    \node (c)at(-1,1){$\delta_b$};
    \draw [-](node cs: name=a)--(b);
    \draw [-](node cs: name=a)--(c);
  \end{tikzpicture}
    \caption{$V^c_{\leq 1}(\{a,b\})$}
\end{figure}
\end{exa}

\bibliographystyle{alphaurl}
\bibliography{ref}

\end{document}